\documentclass[conference]{IEEEtran}
\IEEEoverridecommandlockouts

\usepackage{tabularx}
\usepackage{multirow}
\usepackage{booktabs}
\usepackage[table]{xcolor}
\usepackage[normalem]{ulem}
\usepackage{makecell}
\usepackage{tabularx}
\usepackage{array}
\usepackage{pifont}
\usepackage{url}

\usepackage{cite}
\usepackage{amsmath,amssymb,amsfonts}
\usepackage{algorithmic}
\usepackage{graphicx}
\usepackage{textcomp}
\usepackage{xcolor}
\usepackage{graphicx}

\def\BibTeX{{\rm B\kern-.05em{\sc i\kern-.025em b}\kern-.08em
    T\kern-.1667em\lower.7ex\hbox{E}\kern-.125emX}}

\usepackage{eso-pic}
\newcommand{\ieeecopyrightnotice}{
  \copyright~2026 IEEE. Personal use of this material is permitted. Permission from IEEE must be obtained for all other uses, in any current or future media, including reprinting/republishing this material for advertising or promotional purposes, creating new collective works, for resale or redistribution to servers or lists, or reuse of any copyrighted component of this work in other works.
}
\AddToShipoutPictureBG*{
  \AtPageLowerLeft{
    \put(\LenToUnit{\dimexpr(\paperwidth-\textwidth)/2\relax},\LenToUnit{0.3in}){
      \parbox[b]{\textwidth}{\scriptsize\centering\ieeecopyrightnotice}}}}

\begin{document}

\title{When Demonstrations Fail: Diagnosing the Limits of In-Context Learning in Large Audio-Language Models with Progressive Cue Removal}

\author{
  \IEEEauthorblockN{
    Yen-Ting Piao\textsuperscript{*}\quad
    Jay Chiehen Liao\textsuperscript{*}\quad
    Wei-Tang Chien\textsuperscript{*}\quad
    Toshiki Ogimoto\textsuperscript{*}\\
    Shang-Tse Chen\quad
    Yun-Nung Chen\quad
    Chun-Yi Lee\quad
    Shao-Yuan Lo
  }
  \IEEEauthorblockA{\\
    National Taiwan University
    \\
    \texttt{\{r14922010, r13922210, r13922197, r14922163, stchen\}@csie.ntu.edu.tw}\\
    \texttt{y.v.chen@ieee.org\quad \{cylee, sylo\}@csie.ntu.edu.tw}
  }
  \thanks{\textsuperscript{*}Equal contribution.}
}

\maketitle

\begin{abstract}
    While Large Audio-Language Models (LALMs) have been shown to exhibit degraded instruction-following capabilities, their ability to infer task patterns from in-context examples with audio remains understudied. To address this gap, we design a three-stage evaluation pipeline that progressively reduces textual guidance to systematically evaluate LALMs’ in-context learning ability in the audio modality. Evaluating six LALMs across four audio understanding tasks under two output constraint categories, we uncover a consistent asymmetry across LALMs: in-context demonstrations reliably improve format compliance but fail to improve the core task performance. This suggests that LALMs can glean surface-level formatting patterns from demonstrations but may struggle to leverage cross-modal semantic grounding to reliably infer task objectives from examples with audio, highlighting potential limitations in current cross-modal integration. We further probe how demonstrations are used through two complementary analyses, demonstration label shuffling and attention knockout on demonstration spans, both showing that LALMs leverage in-context examples primarily to establish the output label space and format rather than to learn a meaningful input–output correspondence.\footnote{\url{https://github.com/MiuLab/LALM-ICL}}
\end{abstract}


\section{Introduction}

Large Audio-Language Models (LALMs)~\cite{gong2023joint, chu2024qwen2, lu2025desta25Audio, wang-etal-2024-blsp, xu2025qwen25omnitechnicalreport, microsoft2025phi4minitechnicalreportcompact} extend Large Language Models (LLMs) to the auditory domain, supporting applications such as semantic-level speech question answering, spoken dialogue generation, paralinguistic analysis (e.g., emotion or speaker traits), and general non-speech audio event understanding~\cite{yang2025towards, sakura, zhang2023speechgpt, tang2024salmonn}. Despite this progress, recent studies~\cite{lu2025speechifeval, gao-etal-2025-ifeval, li2025isa} reveal that LALMs often exhibit degraded instruction-following ability compared to their text-only counterparts, particularly under explicit output constraints~\cite{lu2025speechifeval}. This gap is hypothesized by these studies to stem from catastrophic forgetting during speech-text integration, suggesting that multimodal adaptation could affect not only modality alignment but also higher-level task inference mechanisms inherited from text pretraining.

In the text-only domain, in-context learning (ICL) enables LLMs to infer latent task structures and output formats through few-shot demonstrations at inference time~\cite{brown2020language, min2022rethinking, dong-etal-2024-survey, chen-etal-2025-icleval}.
Crucially, ICL and instruction-following share a common underlying mechanism: both require the model to infer task objectives and output format constraints from contextual signals, whether those signals are explicit natural language instructions or implicit demonstrations~\cite{pan2023context, xie2022an}. Therefore, evaluating LALMs' ICL behaviors provides a principled approach for probing whether these higher-level task recognition mechanisms generalize to the audio modality after speech–text integration.

However, despite extensive studies of ICL in text-only LLMs~\cite{brown2020language, min2022rethinking, dong-etal-2024-survey, chen-etal-2025-icleval}, prior work in the auditory domain covers only isolated tasks such as automatic speech recognition~\cite{roll-etal-2025-context} or auditory knowledge editing~\cite{yang2025sake}. To the best of our knowledge, there is no systematic study of how LALMs perform ICL from audio-infused demonstrations. Rather than comparing LALMs with their text-only counterparts, our goal is to disentangle the joint influence of acoustic inputs and in-context demonstrations on task inference.

To bridge this gap, we devise a systematic three-stage evaluation pipeline for analyzing LALMs' ICL ability as illustrated in Figure~\ref{fig:overview}.
We design three stages that progressively reduce textual guidance in the in-context demonstrations, which are all task-domain consistent, while keeping the audio inputs and demonstration outputs unchanged. Such a controlled setup isolates the contribution of textual cues and assesses whether LALMs can infer task objectives and generate correctly formatted responses from audio-text-paired examples.

Across six LALMs, we consistently observe a strongly asymmetric ICL effect:
\textbf{in-context demonstrations only improve format compliance but not core speech task performance}, even when examples with chain-of-thought reasoning traces are provided. Notably, the evaluated LALMs with stronger instruction-following ability undergo more degradation in format compliance when explicit constraint instructions are removed, indicating that \textbf{good instruction-following capability does not imply robust ICL deduction ability} from demonstrations alone.

Overall, our key contributions are threefold: (i) We present the first systematic study of how LALMs perform ICL from demonstrations with audio and text, built on a three-stage pipeline that progressively removes textual cues. (ii) Across six LALMs and four audio understanding tasks, we uncover a consistent asymmetry: demonstrations improve format compliance but fail to improve core task performance, even for models with strong instruction-following ability. (iii) Through label-shuffling and attention-knockout analyses, we show that LALMs exploit demonstrations chiefly to fix the output label space and format rather than genuinely learning the audio–label correspondence.

\section{Three-Stage ICL Evaluation Pipeline}

To investigate whether LALMs exhibit ICL behavior, we adopt a subset of Speech-IFEval~\cite{lu2025speechifeval} as our evaluation testbed, a benchmark of audio understanding tasks augmented with verifiable output constraints~\cite{zhou2023instruction, lu2025speechifeval} that enables systematic assessment of format compliance and speech-processing ability under prompts with audio.

\subsection{Audio Understanding Tasks and Format Constraints}

\begin{figure}[t]
    \centering
    \includegraphics[width=0.95\linewidth]{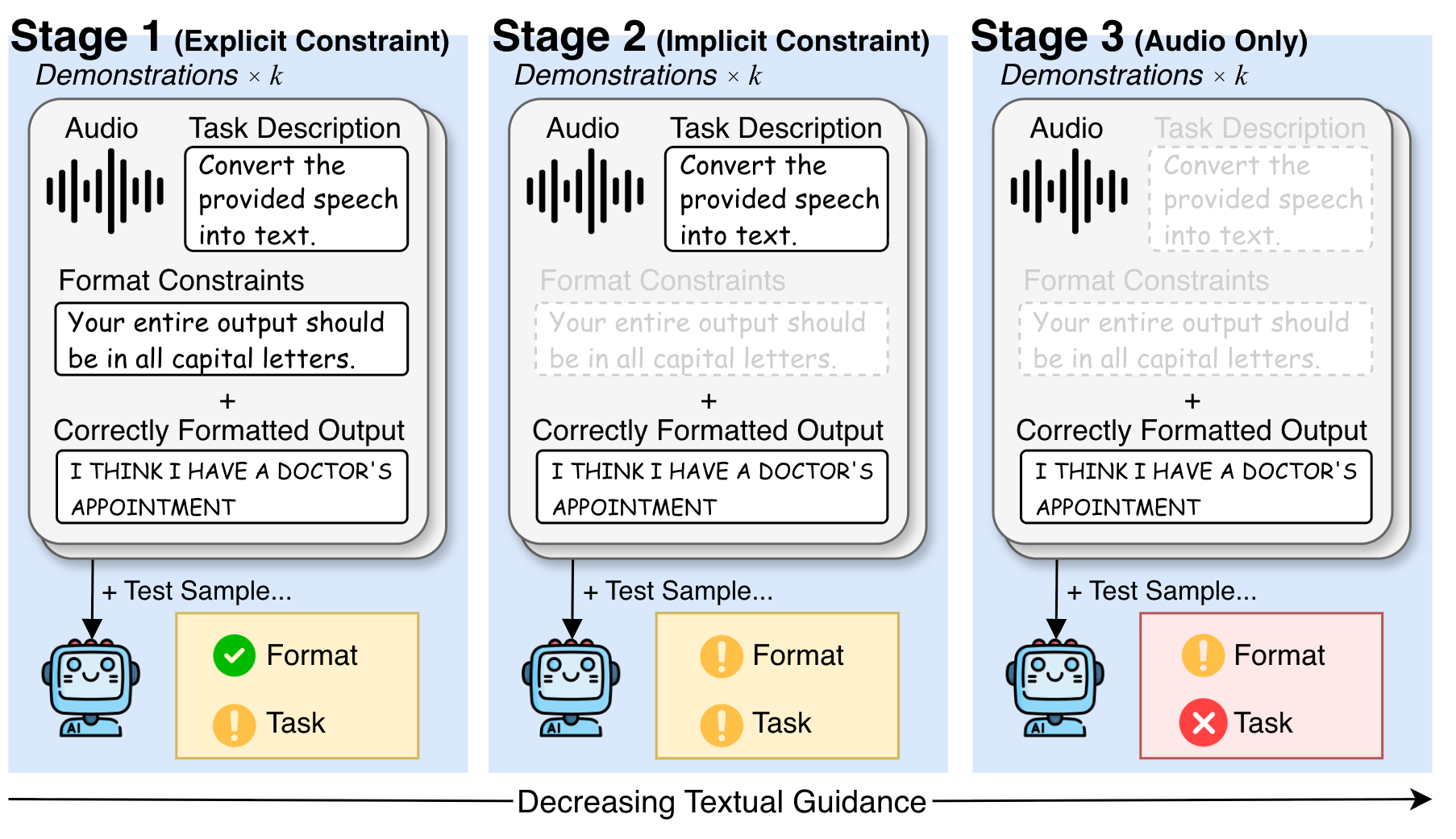}

    \vspace{-7pt}
    \caption{
    Our three evaluation stages: Textual guidance progressively decreases from Stage 1 (Explicit Constraint) to Stage 3 (Audio Only), with removed elements grayed out.
    }
    \label{fig:overview}
    \vspace{-1.5em}
\end{figure}

We adopt the audio understanding tasks from Speech-IFEval, including \textbf{A}utomatic \textbf{S}peech \textbf{R}ecognition (ASR), \textbf{S}peech \textbf{E}motion \textbf{R}ecognition (SER), \textbf{G}ender \textbf{R}ecognition (GR), and \textbf{M}assive \textbf{M}ulti-Task \textbf{A}udio \textbf{U}nderstanding (MMAU)~\cite{sakshi2024mmaumassivemultitaskaudio}.
Also following Speech-IFEval, we consider two constraint categories: \textbf{(1)~Closed-Ended Questions (CEQ)}, which enforce strict deterministic formats (e.g., all-uppercase output, JSON formatting, or a required prefix/suffix); and \textbf{(2)~Chain-of-Thought (CoT)}, which requires the model to produce explicit intermediate reasoning steps alongside the final answer.
Figure~\ref{fig:overview} illustrates an example using the all-uppercase constraint.

\subsection{Construction of ICL Examples}

\textbf{Instruction rephrasing.} For each task-constraint combination, we use GPT-5~\cite{gpt5} to rephrase task instructions and format constraints respectively while preserving semantics, diversifying linguistic expressions across demonstrations and reducing lexical and structural bias.

\textbf{Audio and answer sourcing.} For ASR, SER, and GR, we sample audio from CREMA-D~\cite{6849440}, a corpus disjoint from Speech-IFEval's data containing diverse emotional speech with labelled speaker information.
For MMAU, we sample audio clips from the MMAU dataset, ensuring that none overlap with Speech-IFEval.
To maintain annotation quality, we only include MMAU items with either provided ground-truth answers or confidently human-annotated labels.
Our evaluation dataset contains 988 samples with an average audio duration of 11.97 s, and our demonstration dataset contains 912 samples with an average audio duration of 8.08 s.

\textbf{Format label generation.} CEQ demonstration labels are generated via rule-based transformations on the answers (e.g., uppercasing text), while CoT reasoning traces are generated by Gemini~2.5~Pro~\cite{comanici2025gemini} and verified by a human expert.

The resulting demonstration pool consists of audio-instruction-label triples that are linguistically diverse and disjoint from the evaluation data. They are label-balanced to avoid exposing the models to skewed label distributions during ICL.

\definecolor{basegray}{gray}{0.92}
\newcommand{\base}[1]{\cellcolor{basegray}#1}
\newcommand{\pmcell}[2]{#1\,{\tiny$\pm$\,#2}}
\newcommand{\bestpm}[2]{\textbf{#1}\,{\tiny$\pm$\,#2}}
\newcommand{\worstpm}[2]{\uline{#1}\,{\tiny$\pm$\,#2}}

\begin{table*}[t]
\centering
\footnotesize
\setlength{\tabcolsep}{2pt}
\renewcommand{\arraystretch}{1.15}

\caption{
    \textbf{Format Compliance Rate (FCR, \%) reported with 95\% confidence intervals.}
    ``Few-shot'' is averaged over $k \ge 1$.
    Shaded cells indicate the zero-shot baseline within each task group, and \textbf{bold} marks the highest FCR within each model$\times$task-group block.
}
\label{tab:fcr}

\vspace{-0.8em}

\resizebox{\textwidth}{!}{
\begin{tabular}{@{}l | c c c c c | c c c c c c c @{}}
\toprule
\multirow{4}{*}{\textbf{Model}} &
\multicolumn{5}{c|}{\textbf{Task Group A: ASR/SER/GR/MMAU}} &
\multicolumn{7}{c}{\textbf{Task Group B: ASR/SER/GR}} \\
\cmidrule(lr){2-6}\cmidrule(lr){7-13}

& \multicolumn{1}{c}{\textbf{Zero-shot}} &
  \multicolumn{2}{c}{\textbf{One-shot}} &
  \multicolumn{2}{c|}{\textbf{Few-shot}} &
  \multicolumn{1}{c}{\textbf{Zero-shot}} &
  \multicolumn{3}{c}{\textbf{One-shot}} &
  \multicolumn{3}{c}{\textbf{Few-shot}} \\
\cmidrule(lr){2-2}\cmidrule(lr){3-4}\cmidrule(lr){5-6}
\cmidrule(lr){7-7}\cmidrule(lr){8-10}\cmidrule(lr){11-13}

& \textbf{Stage~1} &
  \textbf{Stage~1} & \textbf{Stage~2} &
  \textbf{Stage~1} & \textbf{Stage~2} &
  \textbf{Stage~1} &
  \textbf{Stage~1} & \textbf{Stage~2} & \textbf{Stage~3} &
  \textbf{Stage~1} & \textbf{Stage~2} & \textbf{Stage~3} \\

\midrule

\textsc{Qwen2-Audio} &
\base{\pmcell{40.08}{3.05}} &
\pmcell{55.67}{3.09} & \pmcell{22.57}{2.60} &
\bestpm{57.06}{3.08} & \pmcell{40.02}{3.01} &
\base{\pmcell{41.61}{3.59}} &
\pmcell{56.17}{3.61} & \pmcell{13.31}{2.48} & \pmcell{27.60}{3.26} &
\bestpm{57.63}{3.60} & \pmcell{32.11}{3.32} & \pmcell{44.40}{3.57} \\

\textsc{DeSTA2.5-Audio} &
\base{\pmcell{95.45}{1.31}} &
\pmcell{98.48}{0.78} & \pmcell{33.81}{2.94} &
\bestpm{98.94}{0.66} & \pmcell{71.38}{2.62} &
\base{\pmcell{95.70}{1.50}} &
\pmcell{98.75}{0.85} & \pmcell{32.87}{3.42} & \pmcell{34.81}{3.47} &
\bestpm{98.79}{0.83} & \pmcell{69.38}{3.15} & \pmcell{67.11}{3.19} \\

\textsc{BLSP-Emo} &
\base{\pmcell{66.80}{2.93}} &
\bestpm{97.98}{0.90} & \pmcell{72.57}{2.78} &
\pmcell{97.48}{0.99} & \pmcell{88.35}{1.93} &
\base{\pmcell{70.60}{3.32}} &
\bestpm{98.20}{1.00} & \pmcell{71.01}{3.30} & \pmcell{60.75}{3.56} &
\pmcell{97.99}{1.05} & \pmcell{90.60}{1.97} & \pmcell{88.12}{2.13} \\

\textsc{Qwen2.5-Omni} &
\base{\pmcell{82.79}{2.35}} &
\pmcell{91.09}{1.78} & \pmcell{58.50}{3.07} &
\bestpm{94.67}{1.40} & \pmcell{82.19}{2.20} &
\base{\pmcell{80.17}{2.91}} &
\pmcell{89.32}{2.26} & \pmcell{54.23}{3.63} & \pmcell{67.96}{3.40} &
\bestpm{93.57}{1.78} & \pmcell{79.66}{2.72} & \pmcell{88.59}{2.13} \\

\textsc{Phi-4-Multimodal} &
\base{\pmcell{60.12}{3.05}} &
\bestpm{98.58}{0.76} & \pmcell{85.12}{2.22} &
\pmcell{98.15}{0.85} & \pmcell{89.85}{1.88} &
\base{\pmcell{53.40}{3.63}} &
\bestpm{98.34}{0.97} & \pmcell{84.19}{2.66} & \pmcell{90.43}{2.15} &
\pmcell{97.76}{1.10} & \pmcell{89.55}{2.22} & \pmcell{96.64}{1.28} \\

\textsc{Gemini 2.5 Flash (On)} &
\base{\pmcell{98.68}{0.73}} &
\pmcell{99.70}{0.39} & \pmcell{69.33}{2.87} &
\bestpm{99.85}{0.30} & \pmcell{78.77}{2.53} &
\base{\pmcell{99.03}{0.76}} &
\pmcell{99.72}{0.46} & \pmcell{72.26}{3.26} & \pmcell{82.39}{2.78} &
\bestpm{99.93}{0.32} & \pmcell{81.14}{2.84} & \pmcell{85.75}{2.55} \\

\textsc{Gemini 2.5 Flash (Off)} &
\base{\pmcell{91.90}{1.71}} &
\pmcell{99.19}{0.59} & \pmcell{72.98}{2.77} &
\bestpm{99.48}{0.48} & \pmcell{80.68}{2.45} &
\base{\pmcell{91.26}{2.07}} &
\pmcell{98.89}{0.81} & \pmcell{70.18}{3.33} & \pmcell{80.03}{2.91} &
\bestpm{99.29}{0.65} & \pmcell{77.79}{3.02} & \pmcell{83.03}{2.74} \\

\bottomrule
\end{tabular}
}

\vspace{-1.5em}
\end{table*}

\begin{figure*}
    \centering
    \vspace{2mm}
    \includegraphics[width=\linewidth]{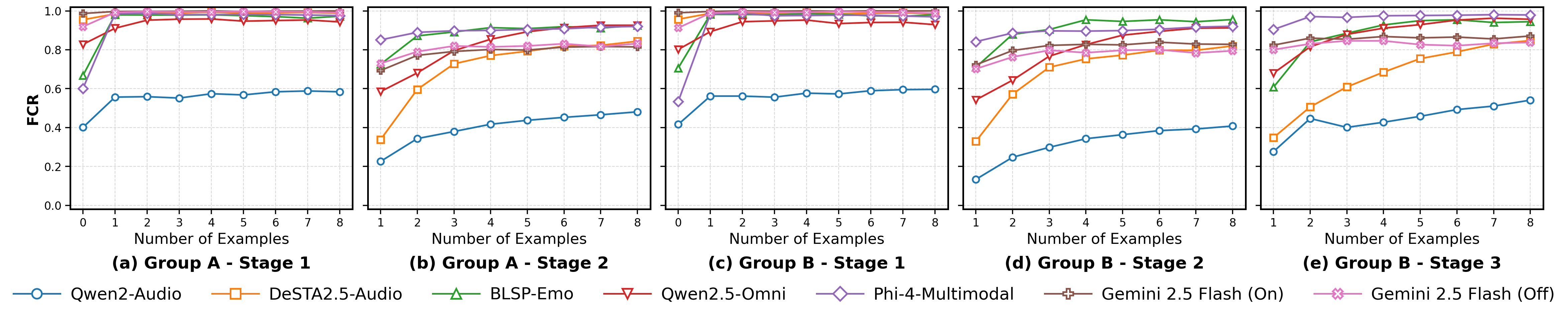}
    \vspace{-8mm}
    \caption{
        \textbf{Format Compliance Rate (FCR) of the Number of In-Context Examples $k$ across the evaluation stages for Groups A and B.} This resolves the aggregated ``Few-shot'' columns of Table~\ref{tab:fcr} into the full per-shot trajectory. Please note that zero-shot ($k=0$) works for Stage 1 only.
    }
  \label{fig:fcr}
  \vspace{-10pt}
\end{figure*}

\subsection{Three-Stage Evaluation Pipeline}
\label{sec:three_stages}

The core evaluation methodology is illustrated in Figure~\ref{fig:overview}.
We design a three-stage pipeline that progressively reduces the amount of explicit textual instruction provided, thereby probing the degree to which LALMs can learn from audio examples when textual guidance is systematically withheld.
This design allows us to attribute any observed ICL effect to a specific type of contextual signal.

In \textbf{Stage 1 (Explicit Constraint)}, each demonstration example and test sample contains both a task description and explicit format constraint instruction alongside the audio input. Demonstrations are drawn from the same task-constraint category as the test sample and always contain correctly formatted outputs. This stage serves as a strong upper-bound condition: textual instructions fully specify both the task objective and how the output should be formatted, so any observed ICL effect reflects learning on top of already complete textual guidance.

In \textbf{Stage 2 (Implicit Constraint)}, the explicit format constraint instructions are removed from both the demonstration examples and test samples while other parts remain intact. This stage tests if the models can infer formatting conventions from example outputs alone, without being explicitly told the rules.

In \textbf{Stage 3 (Audio Only)}, task descriptions are further removed, providing the models with only audio inputs paired with correctly formatted outputs as demonstrations. This most challenging stage requires the models to jointly infer task objectives and formatting patterns from the relationship between audio inputs and textual outputs across demonstrations, constituting a strict test of cross-modal ICL. Note that MMAU is excluded at this stage as its textual instruction is necessary to answer the question, making the task ill-defined without it.

\section{Experimental Setup}

We explore the effects of the number of in-context examples~($k$) by varying $k$ from $1$ to $8$, a range sufficient to observe convergence trends while remaining computationally feasible. Examples are sampled from the demonstration pool to match the test sample's task domain and constraint category, then inserted into the dialogue context in random order. To anchor the comparison baseline against ICL, Stage 1 includes a zero-shot setting in which the LALMs rely solely on natural language instructions without demonstrations. We omit zero-shot conditions in Stages 2 and 3: format compliance in Stage 2 is undefined without explicit constraints, and providing only audio inputs in Stage 3 renders the task ill-defined.

\subsection{Models}
Our goal is to investigate whether LALMs can integrate auditory information during ICL, but given the heterogeneity of the underlying backbones (e.g., pretrained versus instruction-tuned) across different LALMs, directly comparing them with their text-only counterparts lacks strict experimental control. Therefore, we focus our comparative design exclusively within the audio-conditioned setting, rather than evaluating their retention of text-based ICL abilities.
We evaluate five open-source LALMs of various sizes: Qwen2-Audio (8.4B)~\cite{chu2024qwen2}, DeSTA2.5-Audio (9.7B)~\cite{lu2025desta25Audio}, BLSP-Emo (8.5B)~\cite{wang-etal-2024-blsp}, Qwen2.5-Omni (10.7B)~\cite{xu2025qwen25omnitechnicalreport}, and Phi-4-Multimodal (5.6B)~\cite{microsoft2025phi4minitechnicalreportcompact}. In addition, we include one proprietary model, Gemini 2.5 Flash~\cite{comanici2025gemini}, evaluated under two configurations: dynamic thinking enabled (``On'') and thinking entirely disabled (``Off'').

\subsection{Evaluation Metrics}
We evaluate ICL along two complementary dimensions:

\begin{itemize}
    \item \textbf{Format Compliance Rate (FCR)}: This measures whether a model's response adheres to the prescribed output constraints shown in the instructions or demonstrations. For CEQ constraints, we employ a rule-based processor~\cite{zhou2023instruction}; for CoT, we adopt LLM-as-a-judge~\cite{chiang-lee-2023-large} with GPT-5 mini\footnote{gpt-5-mini-2025-08-07, minimal reasoning, random seed = 42} to determine whether the response exhibits reasoning behavior.

    \item \textbf{Task Performance}: We report standard metrics for each audio understanding task: Word Error Rate (WER; $\downarrow$) for ASR, and accuracy ($\uparrow$) for SER, GR, and MMAU.
    Model responses are preprocessed prior to scoring. CEQ predictions are extracted using a rule-based processor~\cite{lu2025speechifeval}. For CoT-based ASR, instances of hallucinated or non-terminating reasoning~\cite{frieske2024hallucinations, kuan2025can, kumar2025performance} are detected via regular expressions, and the resulting transcriptions are replaced with empty strings; otherwise, GPT-5 mini extracts the transcription from the model's response. For other CoT-based tasks, GPT-5 mini directly determines prediction correctness.
\end{itemize}

To validate the reliability of the LLM judge, we conduct human verification on 300 randomly selected samples, achieving 99\% agreement, indicating strong robustness.

\section{Experimental Results}\label{sec:results}

Table~\ref{tab:fcr} reports the FCR of zero-shot, one-shot, and few-shot across the three stages of our evaluation, with Figure~\ref{fig:fcr} showing each shot's fine-grained results. Since MMAU is not applicable to Stage 3, as described in Sec.~\ref{sec:three_stages}, we present two task groups: Task Group A covers all audio understanding tasks for Stages 1--2, while Task Group B excludes MMAU and covers Stages 1--3. Figures~\ref{fig:task_perf_ceq} and \ref{fig:task_perf_cot} show the trends in task performance across varying numbers of in-context examples for the three stages under CEQ and CoT constraint categories, respectively.

\subsection{Impact of Demonstrations with Explicit Format Constraints}

\textbf{FCR is near-saturated for instruction-strong models, yet one-shot still yields large compliance jumps for others.}
As shown in Table~\ref{tab:fcr}, Gemini 2.5 Flash (On) is essentially at the ceiling even in Stage~1 zero-shot (Task Group A: 98.68\%; Task Group B: 99.03\%), and DeSTA2.5-Audio is likewise highly compliant (A: 95.45\%; B: 95.70\%).
In contrast, several models are only moderately compliant at zero-shot but jump to near-ceiling with a single example (e.g., BLSP-Emo: 66.80\%$\rightarrow$97.98\%, Phi-4-Multimodal: 60.12\%$\rightarrow$98.58\% in Task Group A), suggesting that even with explicit constraints, demonstrations help ``instantiate'' the intended output schema. Qwen2-Audio is the clear outlier, staying low even at few-shot (A: 57.06\%; B: 57.63\%), indicating difficulty in consistently adhering to output constraints.

\textbf{Task performance is mostly stable across the number of examples.}
In (a1)–(d1) of Figures~\ref{fig:task_perf_ceq} and~\ref{fig:task_perf_cot}, most curves are relatively flat for CEQ and fluctuate for CoT, consistent with Stage~1 being a near-upper-bound setting where text already specifies the task and format.
WER for ASR is generally low and MMAU primarily reflects intrinsic model capabilities and exhibits weak $k$-dependence.

\textbf{High format compliance is independent of high task performance.}
Table~\ref{tab:fcr} and Figures~\ref{fig:task_perf_ceq}--\ref{fig:task_perf_cot} jointly suggest that all LALMs benefit from ICL examples to infer output format, yet they cannot leverage such examples to internalize the core concepts of accomplishing speech-processing tasks, even under CoT, where models are provided with reasoning traces generated by a stronger LALM.
This phenomenon also persists in other stages;
a detailed discussion is presented in Sec.~\ref{sec:result_summary}.

\subsection{Impact of Removing Explicit Format Constraints}

\textbf{Format inference from demonstrations alone is unreliable for LALMs.}
Comparing Stages 1 and 2 in Table~\ref{tab:fcr}, FCR consistently decreases across all models, with few-shot drops ranging from 7.39\% (BLSP-Emo: 97.99\% $\rightarrow$ 90.60\% for Task Group B) to 29.41\% (DeSTA2.5-Audio: 98.79\% $\rightarrow$ 69.38\% for Task Group B). This indicates that without explicit constraint instructions, format inference from demonstrations remains challenging for LALMs.
Furthermore, in Task Group A, LALMs achieving the highest Stage 1 zero-shot FCR show substantial FCR drops in Stage 2, namely DeSTA2.5-Audio (95.45\% $\rightarrow$ 71.38\%) and two Gemini variants (On: 98.68\% $\rightarrow$ 78.77\%; Off: 91.90\% $\rightarrow$  80.68\%). This reveals that strong instruction-following ability \emph{does not guarantee} robust format deduction.

\begin{figure}
    \centering
    \includegraphics[width=0.95\linewidth]{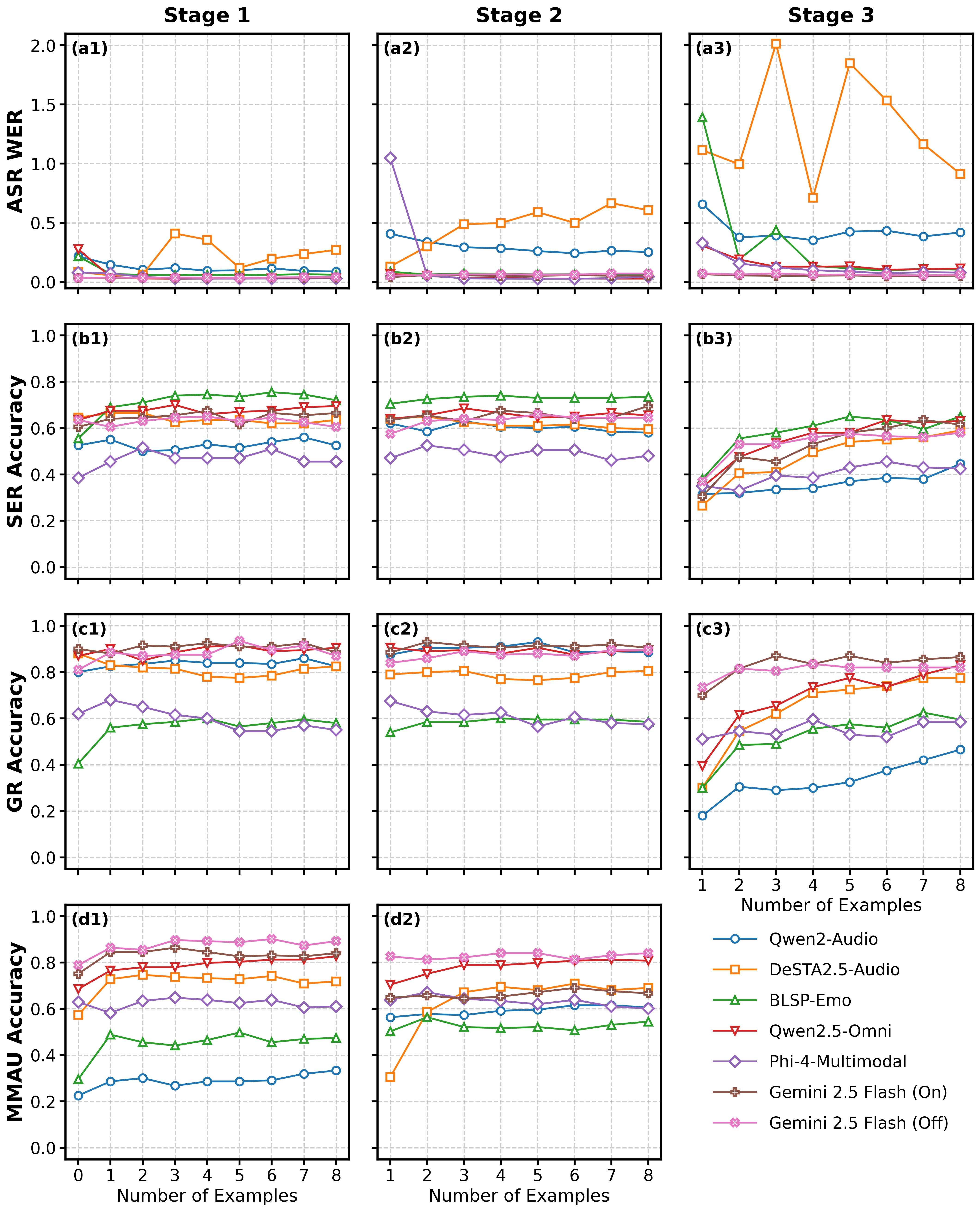}
    \vspace{-1em}
    \caption{
        \textbf{Task Performance of CEQ.}
        Performance vs.\ number of in-context examples $k$ under the three-stage evaluation pipeline for closed-ended questions.
        Rows correspond to tasks (ASR in WER$\downarrow$; SER/GR/MMAU in accuracy$\uparrow$), and columns correspond to Stages~1--3.
    }
  \label{fig:task_perf_ceq}
  \vspace{-10pt}
\end{figure}

\textbf{Task performance remains stable despite constraint changes.}
Unlike FCR, removing explicit output constraints does not lead to a systematic change in task performance, as the task description is preserved (see the middle-column sub-figures, (a2)–(d2), of Figures~\ref{fig:task_perf_ceq} and~\ref{fig:task_perf_cot}). As in Stage 1, increasing the number of ICL examples yields only negligible performance fluctuations.

\subsection{Impact of Further Removing Textual Task Descriptions}

\textbf{Removing task descriptions can lead to higher format compliance.}
Comparing the FCR of Stages 2 and 3 for Task Group B in Table~\ref{tab:fcr}, we observe a counterintuitive trend: Most models maintain or even improve FCR in both one-shot and few-shot settings. In few-shot settings, minor degradation occurs only for DeSTA2.5-Audio (69.38\% $\rightarrow$ 67.11\%) and BLSP-Emo (90.60\%$\rightarrow$88.12\%), while Phi-4-Multimodal achieves a high FCR of 96.64\%, nearly matching its Stage 1 score (97.76\%).
This suggests that decreasing textual guidance in demonstrations does not always hinder format inference. Since formatting relies on surface-level patterns that can be matched from audio-label pairs without integrating auditory semantics with textual context, the textual task descriptions retained in Stage 2 may instead introduce distracting variability, consistent with the findings that LALMs are sensitive to instruction phrasing~\cite{li2025isa, 11278041}. Removing them in Stage 3 reduces this variability, enabling more reliable pattern extraction.

\begin{figure}
    \centering
    \includegraphics[width=0.95\linewidth]{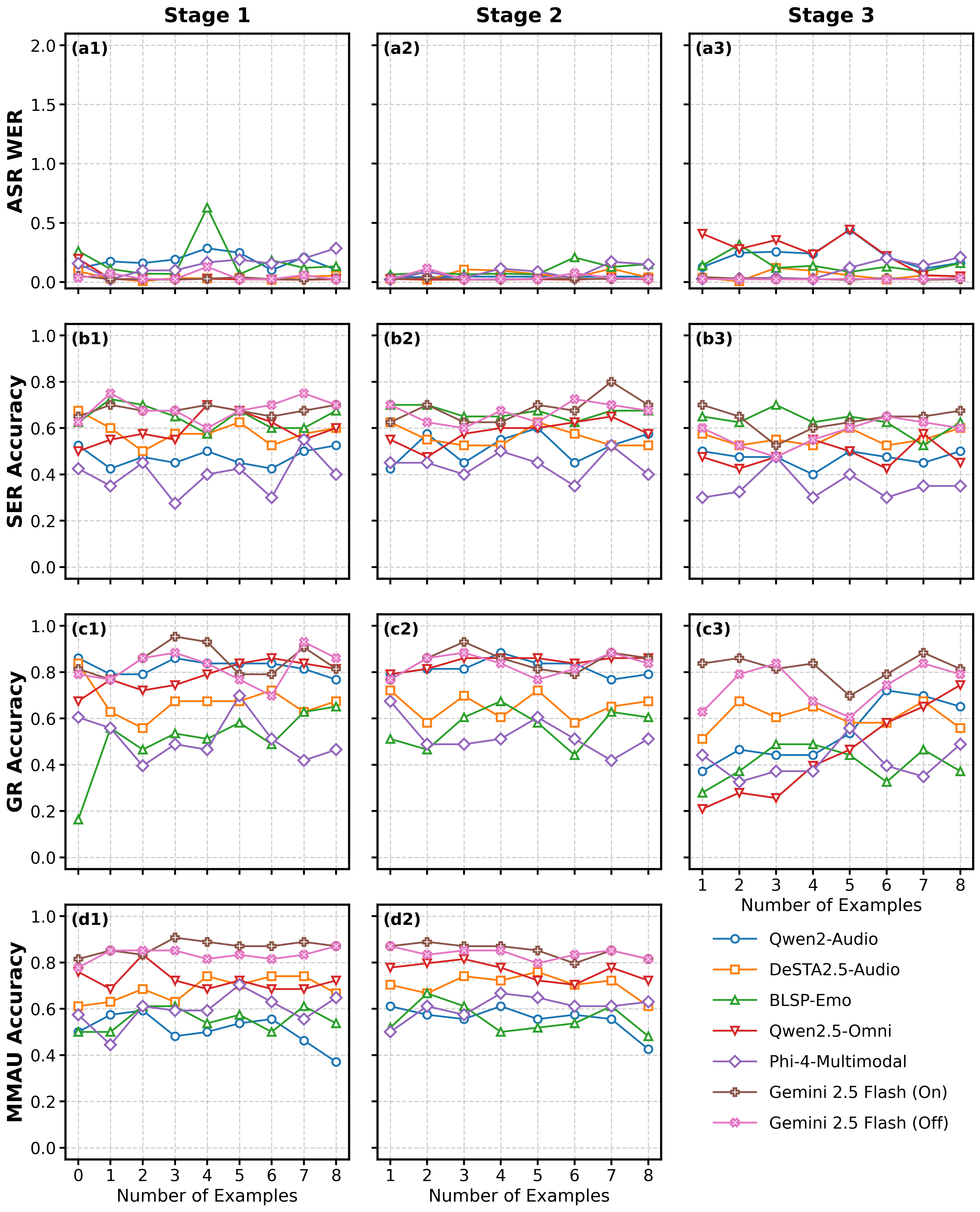}
    \vspace{-1em}
    \caption{
        \textbf{Task Performance of CoT.}
        Performance vs.\ number of in-context examples $k$ under the three-stage evaluation pipeline for chain-of-thought prompting.
        Rows correspond to tasks (ASR in WER$\downarrow$; SER/GR/MMAU in accuracy$\uparrow$), and columns correspond to Stages~1--3.
    }
  \label{fig:task_perf_cot}
  \vspace{-10pt}
\end{figure}

\textbf{Relying solely on audio inputs and formatted outputs to infer the task objective is challenging for LALMs.}
Unlike format patterns, inferring the task objective requires cross-modal semantic grounding that cannot be achieved through surface-level pattern matching alone. As shown in (a3)--(c3) of Figure~\ref{fig:task_perf_ceq}, CEQ task performance in Stage 3 is worse than Stage 2, even with multiple shots, with the gap especially pronounced for ASR in DeSTA2.5-Audio (due to its limited ability to infer the task objective) and GR in Qwen2-Audio.
In contrast, Figure~\ref{fig:task_perf_cot} shows that CoT Stage 3 performance remains close to Stage 2, suggesting that reasoning traces may provide implicit task guidance by exposing the decision process linking audio to text outputs. Notably, Qwen2-Audio and Qwen2.5-Omni exhibit an overall increasing trend in GR accuracy across Stage 3 under CoT, further highlighting this role. Crucially, however, this gain reflects LALM's reliance on textual content within reasoning traces rather than true speech-text integration~\cite{sakura, wang-etal-2025-audio}, as evidenced by the
failure under CEQ where no such textual guidance exists.

\subsection{Summary Across Three Stages}
\label{sec:result_summary}

Our three-stage results suggest that format and task inference rely on fundamentally distinct mechanisms. Format inference requires only surface-level pattern matching, allowing LALMs to maintain or improve FCR in Stage 3 using audio-label pairs alone. Conversely, task inference requires cross-modal semantic grounding: the model must integrate auditory information to map it to a correct response, a capability that remains challenging for current LALMs.
This distinction also clarifies the CoT Stage 3 findings: the ASR performance recovery reflects reliance on textual clues in reasoning traces to infer the task objective rather than genuine speech-text integration, evidenced by the persistent failure of ASR under CEQ. The fact that providing reasoning traces generated by stronger models does not improve task performance in Stages 1 and 2 further confirms that LALMs' ICL limitations are not attributable to insufficient reasoning capacity, but to their \textbf{limited ability to ground auditory perception into task-relevant semantics and integrate it with textual context}, consistent with prior findings on speech/audio reasoning~\cite{sakura}.

\section{Further Analysis: The Role of In-Context Demonstrations}

Our main results show that under the most demanding setting, CEQ in Stage 3, where no textual cues are given, LALMs largely fail to infer the underlying task from demonstrations, with core task performance staying below Stage 2 for most tasks and models even as more examples are added. This failure to learn the input–output task mapping raises a central question about \emph{how} demonstrations are used at all: do they help by conveying the audio–label correspondence, or only by conveying the label space and output format? To answer this, we probe the demonstrations from two complementary angles, both under the Stage 3 CEQ setting.
First, at the \emph{input} level, we break the audio–label correspondence while preserving the label set and format, and observe whether task performance is affected (Sec.~\ref{subsec:label_shuffling}).
Second, at the \emph{internal} level, we use attention knockout to selectively block the model's access to the demonstration audio or the demonstration labels, isolating which span the prediction actually depends on (Sec.~\ref{subsec:attn_knockout}).
Both analyses use Qwen2.5-Omni and Phi-4-Multimodal, two of the models with the best FCRs in our main experiments.

\subsection{Demonstration Label Shuffling}
\label{subsec:label_shuffling}

Following label-perturbation studies in text-only ICL~\cite{min2022rethinking}, we randomly shuffle the labels across the pool of demonstration examples so that each audio clip is paired with another
example's label. This keeps the label space and output format intact while removing the correct audio-to-label mapping, letting us test whether the model relies on the audio–label correspondence or only
on the label space and format.

\begin{figure}
    \centering
    \includegraphics[width=\linewidth]{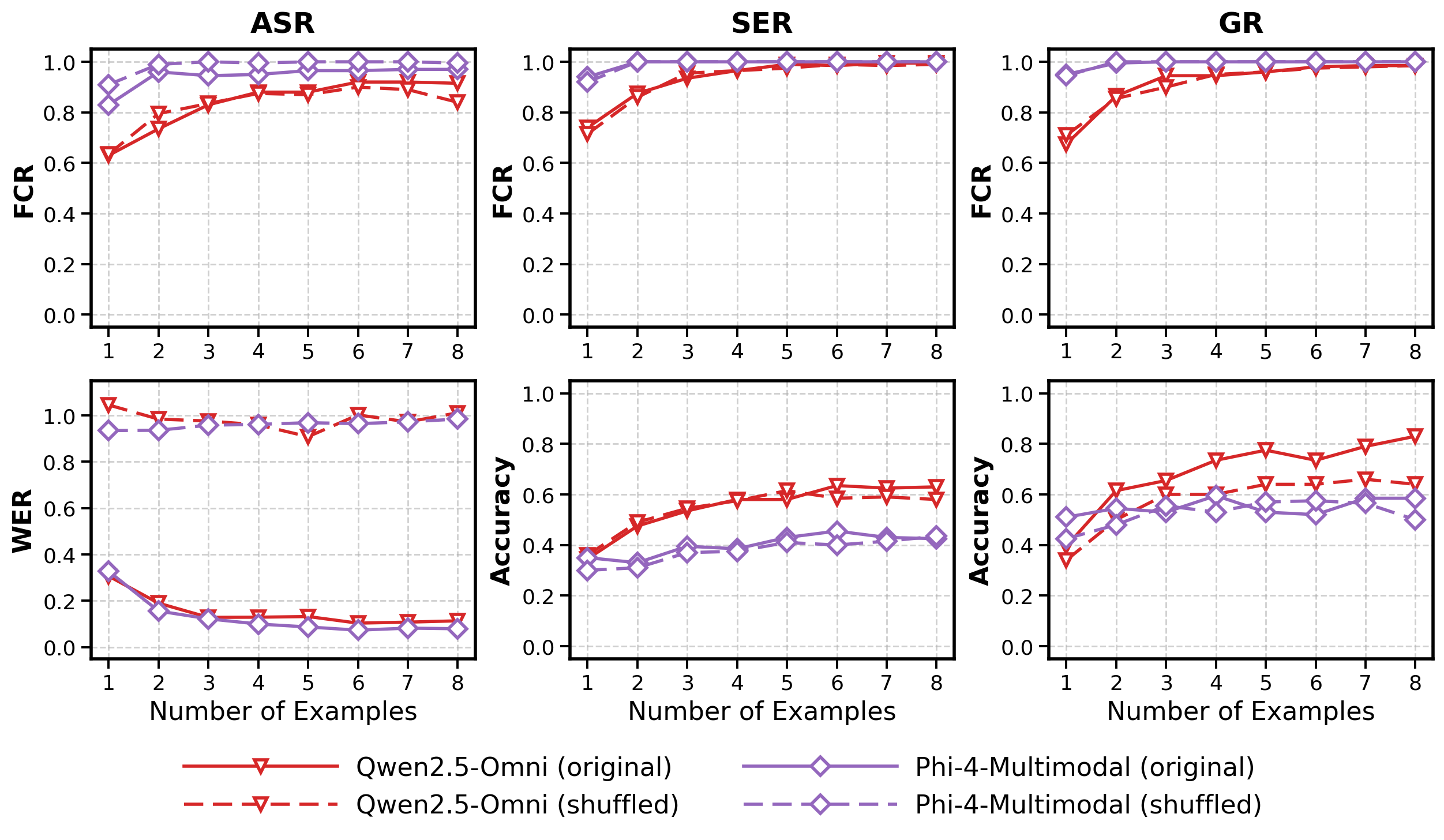}
    \vspace{-2.25em}
    \caption{
        \textbf{Format Compliance Rate (FCR) and Task Performance of Qwen2.5-Omni and Phi-4-Multimodal in Stage~3 (Audio-Only) with Original and Shuffled Labels.} WER$\downarrow$; accuracy$\uparrow$.
    }
  \label{fig:task_perf_random_labels}
  \vspace{-10pt}
\end{figure}

Figure~\ref{fig:task_perf_random_labels} reports Stage~3 FCR and task performance curves for the two models under the shuffled-pairing condition relative to the unmodified baseline. Shuffling the audio–label pairs does not affect the format compliance rate in all core tasks, indicating that models can learn output format information from demonstration examples regardless of whether the audio and label are correctly matched. ASR performance degrades noticeably compared to the original Stage 3 results, whereas SER performance remains comparable and GR degrades only for Qwen2.5-Omni at larger $k$. We attribute the degradation in ASR to its diverse and open-ended label space, which is highly dependent on the audio content; without a valid audio–label correspondence, the model can only extract format information from the ICL examples but cannot infer the correct transcription. In contrast, SER and GR have compact label spaces (four and two categories, respectively) allowing the model to establish both the label space and output format from the label text alone, largely without relying on the paired audio. This suggests that LALMs mostly do not learn input–output correspondence from ICL examples, but instead use demonstrations primarily to establish the label space and output format.

\subsection{Attention Knockout on Demonstration Audio and Labels}
\label{subsec:attn_knockout}

\begin{figure}
    \centering
    \includegraphics[width=\linewidth]{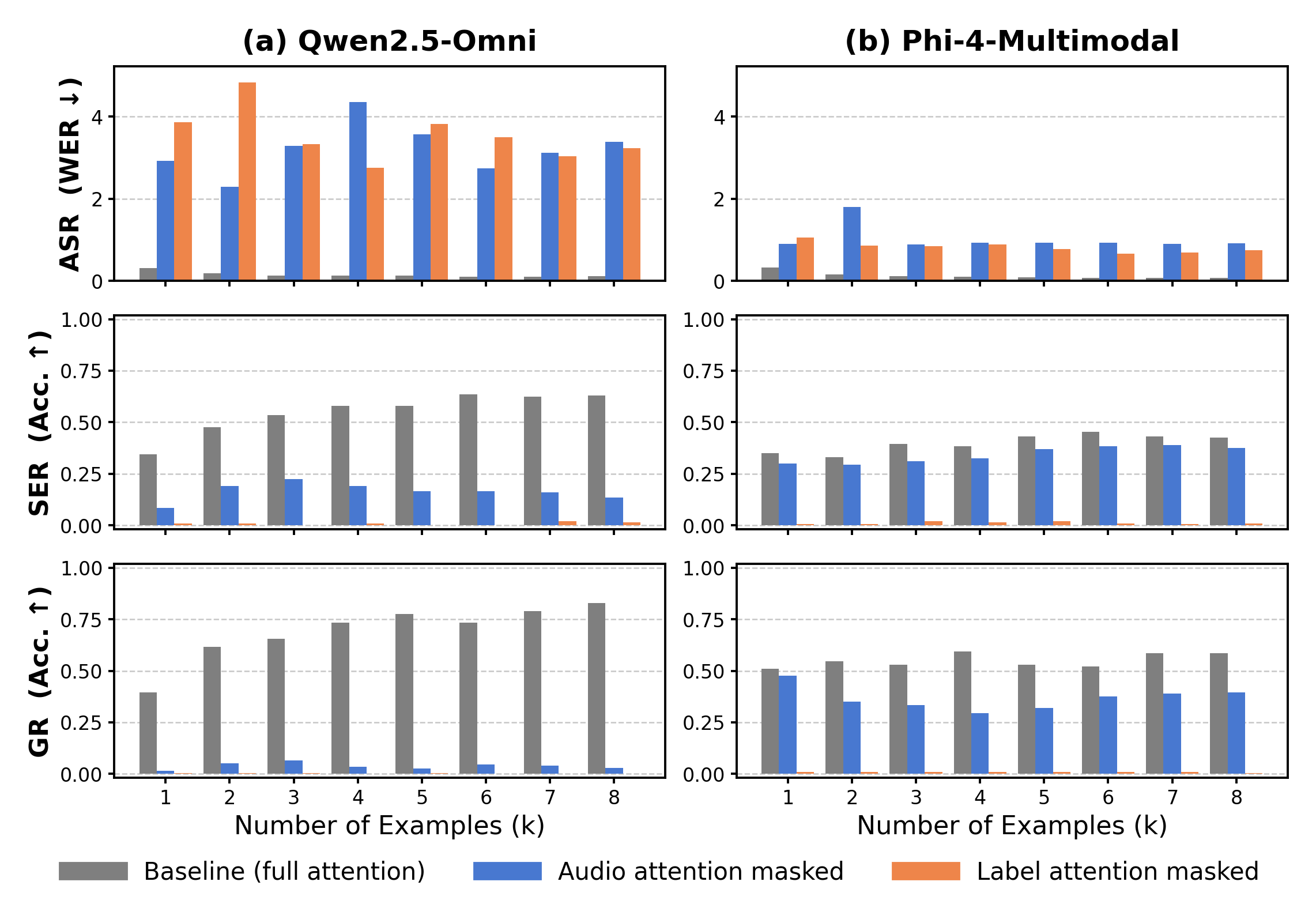}
    \vspace{-2.75em}
    \caption{\textbf{Stage 3 task performance under attention knockout}, comparing the baseline (gray) against masking attention to the demonstration audio (blue) or label (orange) span across 1–8 shots  (WER$\downarrow$; accuracy$\uparrow$).}
  \label{fig:task_performance_attention_knockout}
  \vspace{-10pt}
\end{figure}

To further verify that LALMs use the demonstrations primarily to establish the label space and output format, rather than to learn a genuine audio–label correspondence, we apply attention knockout~\cite{geva-etal-2023-dissecting, 11092754, 11094641, Selvakumar_2026_CVPR}, an interpretability technique that severs information flow by blocking attention to a chosen set of positions. In the Stage~3 prompt, each demonstration consists of an audio span and a label span. To knock out a span, we add a large negative bias ($-\infty$) to the additive attention mask at its key positions, so that every query, at every layer and every decoding step, is prevented from attending to it; the rotary position indices are left untouched so that the remaining tokens retain their original positions. We knock out either the audio span or the label span across all demonstrations in the prompt, and compare the resulting task performance against the unmodified baseline.

Figure~\ref{fig:task_performance_attention_knockout} reports Stage~3 task performance for the two models when attention to the demonstration audio or label span is knocked out, relative to the unmodified
baseline. The effect is strongly asymmetric across tasks and spans. For ASR, both knockouts raise WER far above the low baseline, with the two spans comparably damaging; unlike the closed, low-cardinality
label sets of the classification tasks, ASR has an open-ended label space, so the model cannot fall back on the demonstration labels alone and removing either span severely disrupts transcription. On the
classification tasks, the picture is sharper: masking the demonstration labels drives both SER and GR accuracy to near zero for both models, whereas masking the demonstration audio leaves accuracy
clearly above zero rather than collapsing it (a larger drop but still well above zero for Qwen2.5-Omni, and only a modest drop for Phi-4-Multimodal), the sole exception being GR on Qwen2.5-Omni, where
audio knockout also collapses accuracy. This label-dominant pattern is consistent with our earlier finding that LALMs rely on the demonstration labels to establish the output label space and format, while
the strong ASR degradation and the Qwen2.5-Omni GR collapse under audio knockout indicate that the demonstration audio is not entirely ignored.

\section{Conclusion and Limitations}

Through a three-stage evaluation pipeline that progressively removes textual guidance, we observe an asymmetry in ICL ability across six LALMs and four audio understanding tasks: in-context demonstrations consistently boost format compliance yet do not improve core task performance.
Further analyses through shuffling of demonstration audio–label pairings and attention knockout provide evidence that predictions depend predominantly on the demonstration labels rather than the audio.
These findings suggest a need for better cross-modal integration training paradigms for LALMs.
Nevertheless, our evaluation focuses on format-constrained audio understanding tasks; extending the evaluation to open-ended generation, richer semantic settings, more diverse tasks such as speech synthesis, and investigating demonstration selection and ordering
remains an important direction for future work.

\section*{Acknowledgements}
This work was supported by Taiwan's National Science and Technology Council (NSTC) and the Ministry of Education (MOE)'s Higher Education Sprout Project (Featured Area Research Center Program) under Grants 115-2634-F-002-012, 112-2223-E-002-012-MY5, 115-2628-E-002-023-MY4, 114-2221-E-002-069-MY3, 113-2221-E-002-212-MY3, and 115L900901. This work was also supported by the NTU AI Center of Research Excellence (AI-CoRE). Shao-Yuan Lo is honored to be supported by the MOE's Yushan Young Fellow Program. We also thank the National Center for High-performance Computing (NCHC) of NARLabs for providing computational and storage resources.

In this work, generative AI tools, including LLMs, were used for dataset construction, as judge models for evaluation, and for auxiliary purposes only, namely coding assistance and language editing (e.g., grammar and style refinement).

\bibliographystyle{IEEEtran}
\bibliography{mybib}

\end{document}